\documentclass[submission, creativecommons, sharealike, noncommercial]{eptcs}
\usepackage[utf8]{inputenc}
\usepackage[english]{babel}
\usepackage[T1]{fontenc}
\usepackage{amsmath}
\usepackage{amsthm}
\usepackage{amssymb}
\usepackage{amsfonts}
\usepackage{tabularx}
\usepackage{multicol}
\usepackage{bussproofs}
\EnableBpAbbreviations
\usepackage{txfonts}
\usepackage{pxfonts}
\usepackage{breakcites}
\usepackage{tikz}
\usepackage{fourier}
\usepackage{cmll}
\usepackage[figure]{hypcap}
\usepgflibrary{arrows}
\usepackage{longtable}
\usepackage{xspace}
\usepackage{mathrsfs}

\theoremstyle{plain}
\newtheorem{thm}{Theorem}
\newtheorem{lem}{Lemma}

\newtheorem{fac}{Fact}

\theoremstyle{definition}
\newtheorem{defi}{Definition}
\newtheorem*{proo}{Proof}

\theoremstyle{remark}
\newtheorem{rem}{Remark}
\newtheorem{exes}{Examples}

\hypersetup{
pdfauthor   = {Clément Aubert},
pdftitle    = {Sublogarithmic uniform Boolean proof nets},
pdfsubject  = {Submitted to DICE 2011},
pdfkeywords = {Linear Logic, Boolean circuits, implicit complexity, Boolean proof nets, uniformity, constant-depth circuits, MLLu, mBN, AC, NC, PCC},
pdfcreator  = {PDFLaTeX},
pdfproducer = {PDFLaTeX},
}

\begin{document}
\tikzset{every picture/.style={line width=0.6pt}}
\tikzstyle{every node}=[inner sep=2pt]

\renewcommand{\thefigure}{\arabic{figure}}
\renewcommand{\thetable}{\arabic{table}}
\renewcommand{\labelitemi}{\textendash}

\title{Sublogarithmic uniform Boolean proof nets}
\author{Clément Aubert\thanks{Work partially supported by the French project Complice (ANR-08-BLANC-0211-01).}
\institute{LIPN -- UMR7030, CNRS -- Université Paris 13,\\
99 av. J.-B. Clément, 93430 Villetaneuse, France\\}}
\def\titlerunning{Sublogarithmic uniform Boolean proof nets}
\def\authorrunning{Clément Aubert}
\providecommand{\event}{DICE 2011}
\maketitle

\begin{abstract}
\textbf{Abstract} Using a proofs-as-programs correspondence, Terui was able to compare two models of parallel computation: Boolean circuits and proof nets for multiplicative linear logic. Mogbil \textit{et. al.} gave a logspace translation allowing us to compare their computational power as uniform complexity classes. This paper presents a novel translation in $AC^0$ and focuses on a simpler restricted notion of uniform Boolean proof nets. We can then encode constant-depth circuits and compare complexity classes below logspace, which were out of reach with the previous translations.
\end{abstract}

\section*{Introduction}
\par Boolean proof nets were introduced by Terui in \cite{terui04} to study the implicit complexity of \textit{proofs nets for Multiplicative Linear Logic} \cite{girard96} comparatively to Boolean circuits. Those two models of parallel computation were successfully linked using a \textit{proofs-as-programs} framework, which matches up \textit{cut-elimination} in proof nets with \textit{evaluation} in circuits. Surprisingly \cite{terui04} does not take into account uniformity, which guarantees that the resources needed to build a Boolean circuit is inferior to the computational power it will deliver. \cite{mogbil-uniform} and \cite{mogbil09} studied the Boolean proof nets in a uniform way and introduced some non-de\-termi\-nism in it. As their translation from Boolean circuit families to Boolean proof net families is in logspace ($L$), it remained unknown if the results were still valid when applied to sublogarithmic classes of complexity, that is to say $AC^0$ and $NC^1$. By restricting the Boolean proof nets we use, this paper offers a new proof of the correspondence between circuits and proof nets and extends it to constant-depth circuits.
\par Boolean circuits (\cite{vollmer99}, \autoref{sec1}) and proof nets (\cite{danos89}, \autoref{sec3}) are canonical models of parallel computation, but the latter was mostly seen from the viewpoint of \textit{sequential} implicit complexity. To \textit{evaluate} a proof net is to eliminate its cuts, but to do so with suitable bounds we need to define a parallel elimination. Trying to apply the two usual rules of rewriting ($\to_m$ and $\to_a$) in parallel leads to critical pairs, so we are forced to define a new kind of cut (\textit{tightening-cut}) and a new rewriting rule ($\to_t$). The simulation of this reduction rule by Boolean circuits needs $\textit{UstConn}_2$ gates to be made with efficiency.
\par The proof nets we study are for Multiplicative Linear Logic with unbounded arity (\textbf{MLLu}, \autoref{sec2}) and because of the linearity of this logic, we are forced to keep track of the partial results generated by the evaluation that are unused in the result. Boolean proof nets (\autoref{sec4}) -- as introduced by Terui -- have an expensive way of manipulating this \textit{garbage}. In this paper we introduce \textit{proof circuits} (\autoref{sec5}) as a refinement of the Boolean proof nets that are simpler to manipulate. They are made of \textit{pieces} which translate \textit{gates} and compose easily, so that we reduce the size of the proof net translating the Boolean circuits. In \autoref{sec6} we conclude our paper with our main result (\autoref{NCtoBC}): there exists a constant-depth reduction from Boolean circuit families to proof circuit families. So our new framework offers a variant of the proofs for complexity results and extends them to small classes of complexity, in a uniform way.
\section{Boolean circuits}\label{sec1}
\par Boolean circuits (\autoref{bool_c}) are of great interest in the study of complexity, for instance because of the efficiency of their parallel evaluation. One of their features is that they work only on inputs of fixed length, and that forces us to deal with \textit{families} of Boolean circuits -- and there arises the question of \textit{uniformity} (\autoref{unif1}).
\begin{defi}[Boolean function] A \textit{$n$-ary Boolean function} $f^n$ is a map from $\{ 0, 1 \}^n$ to $\{ 0, 1 \}$. A \textit{Boolean function family} is a sequence $f = (f^n)_{n \in \mathbb{N}}$ and a \textit{basis} is a set of Boolean functions and Boolean function families. We set :
\[\mathfrak{B}_0 = \{ \neg, \vee^2, \wedge^2\} \text{ and } \mathfrak{B}_1 = \{ \neg, (\vee^n)_{n \geqslant 2}, (\wedge^n)_{n \geqslant 2} \}\]
The Boolean function $\textit{UstConn}_2$, given in input the coding of an undirected graph $G$ of degree at most $2$ and two names of gates $s$ and $t$, outputs $1$ iff there is a path between $s$ and $t$ in $G$.
\end{defi}
\begin{defi}[Boolean circuits]\label{bool_c}
 Given a basis $\mathfrak{B}$, a \textit{Boolean circuit over $\mathfrak{B}$ with $n$ inputs} $C$ is a directed acyclic finite and labeled graph. The nodes of fan-in $0$ are called \textit{input nodes} and are labeled with $x_1, \hdots, x_n, 0, 1$. Non-input nodes are called \textit{gates} and each one of them is labeled with a Boolean function from $\mathfrak{B}$ whose arity coincides with the fan-in of the gate. There is a unique node of fan-out $0$ which is the \textit{output gate}. We indicate with a subscript the number of inputs: a Boolean circuit $C$ with $n$ inputs will be named $C_n$.
\par The \textit{depth of a Boolean circuit $C_n$} $d(C_n)$ is the length of the longest path between an input node and the output gate. Its \textit{size} $|C_n|$ is its number of nodes. We will only consider Boolean circuits of size $n^{O(1)}$, that is to say polynomial in the size of their input.
\par $C_n$ \textit{accepts a word} $w \equiv w_1 \hdots w_n \in \{0, 1\}^n$ if $C_n$ evaluates to $1$ when $w_1, \hdots, w_n$ are respectively assigned to $x_1, \hdots, x_n$.
A \textit{family of Boolean circuits} is an infinite sequence $C = (C_n)_{n \in \mathbb{N}}$ of Boolean circuits, $C$ \textit{accepts a language} $X \subseteq \{0, 1\}^*$ iff for all $w \in X$, $C_{|w|}$ accepts $w$.
\end{defi}
\par We now recall the definition of the \textit{Direct Connection Language} of a family of Boolean circuits, an infinite sequence of tuples that describes it totally.
\begin{defi}[Direct Connection Language \cite{vollmer99}]
 Given $\overline{(.)}$ a suitable coding of integers and $C = (C_n)_{n \in \mathbb{N}}$ a family of Boolean circuits over a basis $\mathfrak{B}$, its \textit{Direct Connection Language} -- written $L_{DC}(C)$ -- is the set of tuples $<y, \overline{g}, \overline{p}, \overline{b}>$, such that for $|y| = n$, we have: $g$ is a gate in $C_n$, labeled with $b \in \mathfrak{B}$ if $p = \epsilon$, else $b$ is its $p^{th}$ predecessor.
\end{defi}
\begin{defi}[Uniformity \cite{barrington90}]\label{unif1}
 A family $C$ is said to be \textit{D\-LOG\-TIME}-uniform if there exists a deterministic Turing Machine with random access to the input tape that given $L_{DC}(C)$, $n$ and $\overline{g}$ outputs in time $O(\log(|C_n|))$ any information (position, label or predecessors) about the gate $g$ in $C_n$.
\end{defi}
\par Despite the fact that a \textit{D\-LOG\-TIME} Turing Machine has more computational power than a con\-stant-depth circuit, “\textit{a consensus has developed among researchers in circuit complexity that this \textit{D\-LOG\-TIME} uniformity is the ‘right’ uniformity condition}” for small complexity classes \cite{hesse02}. Any further reference to uniformity is to be read as \textit{D\-LOG\-TIME} uniformity.
\begin{defi}[$AC^i$, $NC^i$]
For all $i \in \mathbb{N}$, given $\mathfrak{B}$ a basis, a language $X \subseteq \{0, 1\}^*$ belongs to the class $AC^i(\mathfrak{B})$ (resp. $NC^i (\mathfrak{B})$) if $X$ is accepted by a uniform family of polynomial-size, $\log^i$-depth Boolean circuits over $\mathfrak{B}_1 \cup \mathfrak{B}$ (resp. $\mathfrak{B}_0 \cup \mathfrak{B}$). We set $AC^i(\emptyset) = AC^i$ and $NC^i(\emptyset) = NC^i$.
\end{defi}
\section{\textbf{MLLu}}\label{sec2}
\par Rather than using Multiplicative Linear Logic (\textbf{MLL}) -- which would force us to compose binary connectives to obtains $n$-ary connectives -- we work with \textbf{MLLu} which differs only on the arities of the connectives but better relates to circuits. We write $\overrightarrow{A}$ (resp. $\overleftarrow{A}$) for an ordered sequence of formulae $A_1, \hdots, A_n$,  (resp. $A_n, \hdots, A_1$).
\begin{defi}[Formulae of \textbf{MLLu}]
Given $\alpha$ a literal and $n \geqslant 2$, formulae of \textbf{MLLu} are:
\[A :: = \alpha ~|~ \alpha^{\bot} ~ | ~ \otimes^n(\overrightarrow{A}) ~|~ \parr^n (\overleftarrow{A})\]
 Duality is defined with respect to De Morgan's law : \begin{eqnarray*}
(A ^{\bot})^{\bot} & \equiv & A \\
(\otimes^n (\overrightarrow{A}))^{\bot} & \equiv & \parr^n (\overleftarrow{A^{\bot}}) \\
(\parr^n(\overleftarrow{A}))^{\bot} & \equiv & \otimes^n (\overrightarrow{A^{\bot}})
\end{eqnarray*}
\par As for the rest of this article, consider that $A$, $B$ and $D$ will refer to \textbf{MLLu} formulae. $A[B / D]$ denotes $A$ where every occurrence of $B$ is replaced by an occurrence of $D$. We write $A[D]$ if $B = \alpha$.
\end{defi}
\begin{defi}[Sequent calculus for \textbf{MLLu}]
A \textit{sequent} of \textbf{MLLu} is of the form $\vdash \Gamma$, where $\Gamma$ is a multiset of formulae. The \textit{inference rules} of \textbf{MLLu} are as follow :
\begin{center}
\begin{tabular}{c c}
\AXC{}
\RightLabel{\textit{ax.}}
\UIC{$\vdash A, A^{\bot}$}
\DP
&
\AXC{$\vdash \Gamma_1, A_1$}
\AXC{$\hdots$}
\AXC{$\vdash \Gamma_n, A_n$}
\RightLabel{$\otimes^n$}
\TIC{$\vdash \Gamma_1, \hdots, \Gamma_n, \otimes^n(\overrightarrow{A})$}
\DP
\\
\\
\AXC{$\vdash \Gamma, A$}
\AXC{$\vdash \Delta, A^{\bot}$}
\RightLabel{\textit{cut}}
\BIC{$\vdash \Gamma, \Delta$}
\DP
&
\AXC{$\vdash \Gamma, \overleftarrow{A}$}
\RightLabel{$\parr^n$}
\UIC{$\vdash \Gamma, \parr^n (\overleftarrow{A})$}
\DP
\end{tabular}
\vspace{1em}
\end{center}
\textit{Derivations of \textbf{MLLu}} are built with respect to those rules. \textbf{MLLu} has neither weakening nor contraction, but admits implicit exchange and eliminates cuts. The formulae $A$ and $A^{\bot}$ in the rule \textit{cut} are called the \textit{cut formulae}.
\end{defi}
\section{Proof nets}\label{sec3}
\par Proof nets are a parallel syntax for \textbf{MLLu} that abstract away everything irrelevant and only keep the structure of the proofs. We introduce measures (\autoref{measures}) on them in order to study their structure and complexity, and a parallel elimination of their cuts (\autoref{pa_elim}).
\begin{defi}[Links]
We introduce in \autoref{links} three sorts of \textit{links} -- $\bullet$, $\otimes^n$ and $\parr^n$ -- which correspond to \textbf{MLLu} rules.
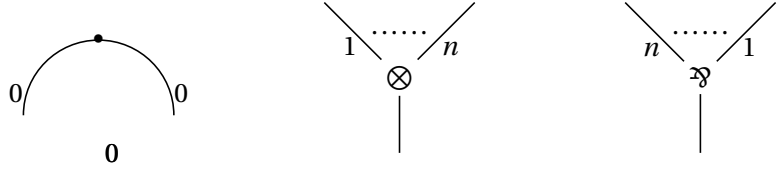
\begin{figure}
\begin{center}
\begin{tikzpicture}
\draw (1,0.5)  arc  (0:180:1);
\draw (0,1.5)  node{$\bullet$};
\node (0) at (-1.1, 0.8) {$0$};
\node (0) at (1.1, 0.8) {$0$};
\node (tenseur) at (4,1) {$\bigotimes$};
\draw (tenseur) to (4, 0) node[right, midway]{$0$};
\draw (tenseur) to node[midway, left, below]{$1~$} (3, 2);
\draw (tenseur) to node[pos=0.2, right]{$~n$} (5, 2); 
\draw (4, 1.6) node{$\hdots \hdots$};
\node (par) at (8, 1) {$\parr$};
\draw (par) to node[pos=0.2, left]{$n~$} (7, 2);
\draw (par) to node[midway, right, below]{$~1$} (9, 2);
\draw (8, 1.6) node{$\hdots \hdots$};
\draw (par) to (8, 0) node[right, midway]{$0$};
\end{tikzpicture}
\caption{$ax$-link, $\otimes^n$-link and $\parr^n$-link} \label{links}
\end{center}
\end{figure}
\par Every link may have two kinds of \textit{ports}: \textit{principal} ones, indexed by $0$ and written below, and \textit{auxiliary} ones, indexed by $1, \hdots, n$ and written above. The auxiliary ports are ordered, but as we always represent the links as in \autoref{links}, we may safely omit the numbering. Axiom links have two principal ports, both indexed with $0$, but we can always differentiate them (for instance by naming them $0_r$ and $0_l$) if we need to.
\end{defi}
\begin{rem}
 There is no sort cut: a cut is represented with an edge between two principal ports.
\end{rem}
\begin{defi}[Decorated derivations and proof net]
Given a derivation of \textbf{MLLu}, we decorate it in the following way:
\begin{itemize}
 \item an index is associated to every formula. The formulae introduced by an axiom have the same proper index, and the formulae introduced by a logical rules ($\otimes^n$ and $\parr^n$) have a \textit{fresh} index,
  \item a \textit{description} is associated to every sequent.
\end{itemize}
\par The rules given in \autoref{decorpn} indicate how a proof is decorated and how to build a \textit{proof net} from a description.
\begin{figure}[t]
\small{
\begin{eqnarray*}
\AXC{}
\RightLabel{$ax.$}
\UIC{$\vdash p : A, p : A^{\bot} \triangleright ax_p$}
\DP
&
\rightsquigarrow
&
\raisebox{-0.5cm}{
\hspace{0.3cm}
\begin{tikzpicture}
\draw (1,0.5)  arc  (0:180:1);
\draw (0,1.5)  node{$\bullet$};
\node (p) at (0.6, 1.5) {$p$};
\node (p') at (-0.6, 1.5) {$p$};
\end{tikzpicture}
}
\\
\\
\AXC{$\{\vdash \Gamma_i, p_i : A_i \triangleright \mathscr{D}(P_i)\}_{1 \leqslant i \leqslant n} $}
\RightLabel{$\otimes^n$}
\UIC{$\vdash \Gamma_1, \hdots, \Gamma_n, s : \otimes^n(\overrightarrow{A}) \triangleright tensor^{p_1, \hdots, p_n}_s (\mathscr{D}(P_1), \hdots, \mathscr{D}(P_n))$}
\DP
&
\rightsquigarrow
&
\raisebox{-2cm}{
\begin{tikzpicture}
\draw (0.5,3) ellipse (10pt and 20pt);
\node (P1) at (0.5, 3) {$P_1$};
\node (p1) at (0.5, 2.5) {$p_1$};
\draw (2.5,3) ellipse (10pt and 20pt);
\node (Pn) at (2.5, 3) {$P_n$};
\node (pn) at (2.5, 2.5) {$p_n$};
\node (tenseur) at (1.5,1) {$\bigotimes$};
\draw (tenseur) -- ++(0, -1) node[pos=0.2, right]{$s$};
\draw (tenseur) to (0.5, 2.3);
\draw (tenseur) to (2.5, 2.3);
\node (points) at (1.5, 3) {$\hdots \hdots$};
\node (points2) at (1.5, 1.6) {$\hdots \hdots$};
\end{tikzpicture}
}
\\
\\
\AXC{$\vdash \Gamma, p : A \triangleright \mathscr{D}(P)$}
\AXC{$\vdash \Delta, q : A^{\bot} \triangleright \mathscr{D}(Q)$}
\RightLabel{$cut$}
\BIC{$\vdash \Gamma, \Delta \triangleright cut^{p, q}(\mathscr{D}(P), \mathscr{D}(Q))$}
\DP
&
\rightsquigarrow
&
\raisebox{-1cm}{
\begin{tikzpicture}
\draw (0.5,3) ellipse (10pt and 20pt);
\node (P) at (0.5, 3) {$P$};
\node (p) at (0.5, 2.5) {$p$};
\draw (2.5,3) ellipse (10pt and 20pt);
\node (Q) at (2.5, 3) {$Q$};
\node (q) at (2.5, 2.5) {$q$};
\draw (0.5, 2.3) .. controls (0.5, 1.5) and (2.5, 1.5) .. (2.5, 2.3);
\end{tikzpicture}
}
\\
\\
\AXC{$\vdash \Gamma, p_n : A_n, \hdots, p_1 : A_1 \triangleright \mathscr{D}(P)$}
\RightLabel{$\parr^n$}
\UIC{$\vdash \Gamma, s : \parr^n(\overleftarrow{A}) \triangleright par^{p_n, \hdots, p_1}_{s} (\mathscr{D}(P))$}
\DP
& \rightsquigarrow
&
\raisebox{-1.5cm}{
\begin{tikzpicture}[scale=0.8]
\draw (1.5,3) ellipse (50pt and 20pt);
\node (par) at (1.5, 1) {$\parr$};
\draw (par) -- ++(0, -1) node[pos=0.2, right]{$s$};
\draw (par) to (0.3, 2.5);
\draw (par) to (2.7, 2.5);
\draw (1.5, 1.6) node{$\hdots \hdots$};
\draw (par) to (1.5, 0);
\node (pn) at (0.2, 2.8) {$p_n$};
\node (points) at (1.5, 2.8) {$\hdots \hdots$};
\node (p1) at (2.8, 2.8) {$p_1$};
\node (P) at (1.5, 3.2) {$P$};
\end{tikzpicture}
}
\\
\end{eqnarray*}
}
Edges representing $\Gamma$ or $\Delta$ are not drawn, $\mathscr{D}(P)$ is one of the description of the proof net $P$.
\caption{From decorated derivations to proof nets}\label{decorpn}
\end{figure}
\par The \textit{type of a proof net $P$} is $\Gamma$ if there exists a decorated derivation of $\vdash \Gamma \vartriangleright \mathscr{D}(P)$: a proof net always has several types, but up to $\alpha$-equivalence (renaming of the literals) we may always assume it has a unique \textit{principal type}. If a proof net may be typed with $\Gamma$, then for every $A$ it may be typed with $\Gamma [A]$. By extension we will use the notion of type of an edge.
\end{defi}
\par The structures obtained by following those rules respect criterion of correctness. For instance two ports of a same link may not be connected, a port may be connected only once and every auxiliary port is connected. We do not have to take into account \textit{pseudo nets}.
\begin{rem}
 The same proof net -- as it abstracts derivations -- may be induced by several descriptions. Conversely, several graphs -- as representations of proof nets -- may correspond to the same proof net: we get round of this difficulty by associating to every proof net a single drawing among the drawings with the minimal number of crossings between edges, for the sake of simplicity. Two graphs representing proof nets that can be obtained from the same description are taken to be equal.
\end{rem}
\begin{defi}[Size and depth of a proof net] \label{measures}
The size $|P|$ of a proof net $P$ is the number of its links. 
\par The depth of a proof net is defined with respect to its type:
\begin{itemize}
 \item The depth of a formula is defined by recurrence:
\begin{description}
\item $d(\alpha) = d(\alpha^{\bot}) = 1$
\item $d(\otimes^n (\overrightarrow{A})) = d(\parr^n (\overleftarrow{A})) = 1 + max(d(A_1), \hdots, d(A_n))$
\end{description}
\item The depth $d(\pi)$ of a derivation $\pi$  is the maximum depth of cut formulae in it.
\item The depth $d(P)$ of a proof net $P$ is
\[min\{d(\pi) ~|~ \pi \text{ can be decorated as } \vdash \Gamma \vartriangleright \mathscr{D}(P) \text{ for some } \Gamma\}\]
The depth $d(P)$ of a proof net depends on its type, but it is minimal when we consider the principal type of the proof net.
\end{itemize}
\end{defi}
\par To make the most of the computational power of proof nets, we need to achieve a speed-up in the number of steps needed to normalize them. If we try roughly to reduce in parallel a cut between two $ax$-links, we are faced with a critical pair. \cite{terui04} avoids this situation by using a \textit{tightening reduction} which eliminates in one step all the cuts between axioms. We can then safely reduce all the other cuts in parallel.
\begin{defi}[Cuts and parallel cut-elimination]\label{pa_elim}
 A cut is an edge between the principal ports of two links. If one of these links is an $ax$-link, two cases occurs:
\begin{description}
\item if the other link is an $ax$-link, we take the maximal chain of $ax$-links connected by their principal ports and defines this set of cuts as a \textit{$t$-cut},
\item otherwise the cut is an \textit{$a$-cut}.
\end{description}
Otherwise it is a \textit{$m$-cut} and we know that for $n \geqslant 2$, one link is a $\otimes^n$-link and the other is a $\parr^n$-link.
\par We define on \autoref{rewriting} three rewriting rules on the proof nets.
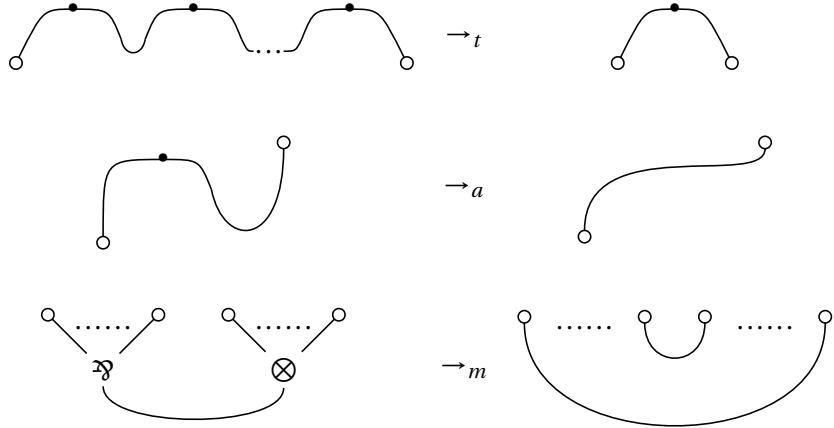
\begin{figure}[t]
For all $\circ \in \{ (\parr^n)_{n \geqslant 2 }, (\otimes^n)_{n \geqslant 2}\}$, $\circ$ may be $\bullet$ in $\to_m$.
\begin{center}
\begin{tikzpicture}[scale=0.8]
\begin{scope}[xshift = -3.5cm, yshift = 0]
\draw [o-] (1, 0) .. controls (1.5, 1) ..  (2, 1);
\node (ax) at (2, 1) {$\bullet$};
\draw (2, 1) .. controls (2.6, 1) .. (2.8, 0.5);
\draw (2.8, 0.5) .. controls (2.85, 0.2) and (3.15, 0.2) .. (3.2, 0.5);
\draw (4, 1) .. controls (3.4, 1) .. (3.2, 0.5);
\draw (4, 1) .. controls (4.6, 1) .. (4.8, 0.5);
\node (ax) at (4, 1) {$\bullet$};
\draw (4.8, 0.5) .. controls (4.9, 0.3) .. (5, 0.3);
\end{scope}
\begin{scope}[xshift=-3.4cm, yshift=0]
\node (points) at (5.15, 0.3) {$\hdots$};
\draw (5.7, 0.5) .. controls (5.6, 0.3) .. (5.4, 0.3);
\draw [-o] (6.5, 1) .. controls (7, 1) ..  (7.5, 0);
\node (ax) at (6.5, 1) {$\bullet$};
\draw (6.5, 1) .. controls (5.9, 1) .. (5.7, 0.5);
\end{scope}
\node (fleche) at (5, 0.5) {$\to_t$};
\draw [o-] (7.5, 0) .. controls (8, 1) ..  (8.5, 1);
\node (ax) at (8.5, 1) {$\bullet$};
\draw [-o] (8.5, 1) .. controls (9, 1) ..  (9.5, 0);
\begin{scope}[xshift=-1cm, yshift=-5cm]
\node (par) at (0, 0) {$\parr$};
\node (ten) at (3, 0) {$\bigotimes$};
\draw[o-] (-1, 1) to (par);
\draw[o-] (1, 1) to (par);
\draw[o-] (2, 1) to (ten);
\draw[o-] (4, 1) to (ten);
\draw (par) .. controls (0, -1) and (3, -1) .. (ten);
\node (points) at (0, 0.7) {$\hdots \hdots$};
\node (points) at (3, 0.7) {$\hdots \hdots$};
\node (fleche) at (6, 0) {$\to_m$};
\begin{scope}[xshift=8cm, yshift=0cm]
\node (points) at (0, 0.7) {$\hdots \hdots$};
\node (points) at (3, 0.7) {$\hdots \hdots$};
\node (par2) at (0, 0) {};
\node (ten2) at (3, 0) {};
\draw[o-o] (-1, 1) .. controls (-1, -1.5) and (4, -1.5) .. (4, 1);
\draw[o-o] (1, 1) .. controls (1, 0) and (2, 0) .. (2, 1);
\end{scope}
\end{scope}
\begin{scope}[xshift=-1cm, yshift=1cm]
\node (ax) at (1, -2.5) {$\bullet$};
\draw (1, -2.5) .. controls (1.6, -2.5) .. (1.8, -3);
\node (p2) at (3, -2) {};
\draw[o-] (0, -4) .. controls (0, -2.5) .. (1, -2.5);
\draw[-o] (1.8, -3) .. controls (2, -4) and (3, -4) .. (p2);
\node (fleche) at (6, -3) {$\to_a$};
\begin{scope}[xshift=8cm, yshift=0cm]
\node (p1) at (0, -4) {};
\node (p2) at (3, -2) {};
\draw [o-o] (p1) .. controls (0, -2) and (3, -3) .. (p2);
\end{scope}
\end{scope}
\end{tikzpicture}
\end{center}
\caption{$t$-, $a$- and $m$-reductions} \label{rewriting}
\end{figure}
For $r \in \{t, a, m\}$, if $Q$ may be obtained from $P$ by erasing all the $r$-cuts of $P$ in parallel, we write $P \rightrightarrows_r Q$. If $P \rightrightarrows_t Q$, $P \rightrightarrows_a Q$ or $P \rightrightarrows_m Q$, we write $P \rightrightarrows Q$. To \textit{normalize a proof net $P$} is to apply $\rightrightarrows$ until we reach a cut-free proof net. $\rightrightarrows^*$ is defined as the transitive reflexive closure of $\rightrightarrows$.
\end{defi}
\begin{thm}[Parallel cut-elimination \cite{terui04}]
Every proof net $P$ normalizes in at most $O(d(P))$ applications of $\rightrightarrows$.
\end{thm}
\par So the time needed to evaluate a proof net is relative to its depth, and can be in the worst case linear in its size -- as for the Boolean circuits.
\section{Boolean proof nets}\label{sec4}
\par In order to compare the complexities of proof nets and of Boolean circuits, we need to define how proof nets represent Boolean values (\autoref{bool_val}) and Boolean functions (\autoref{bool_func}). To study them in a uniform framework we define their Direct Connection Language (\autoref{LDC_pn}).
\begin{defi}[Boolean type, $0$ and $1$ \cite{terui04}] \label{bool_val}
Let $b_0$ and $b_1$ be the two proof nets of type
\[\textbf{B} = \parr^3 (\alpha^{\bot}, \alpha^{\bot}, \alpha \otimes \alpha)\]
respectively used to represent false and true:
\begin{eqnarray*}
\mathscr{D}(b_0)  = par^{q, p, r}_s (tensor^{p, q}_r (ax_p, ax_q)) & b_0  \equiv \raisebox{-0.7cm}{\begin{tikzpicture}
\node (tenb0) at (1,1) {$\bigotimes$};
\node (parb0) at (0,0.5) {$\parr$};
\node (ax1b0) at (0.5,1.2) {$\bullet$};
\node (ax2b0) at (0.5,1.5) {$\bullet$};
\draw (0.5,1.5)  edge[out=5, in=60] (tenb0);
\draw (0.5,1.5) .. controls (-0.8,1.5) .. (parb0);
\draw (0.5,1.2) .. controls (-0.4,1.2) .. (parb0);
\draw (0.5,1.2) to [bend left=10] (tenb0);
\draw (tenb0) to [bend left=10] (parb0);
\draw (0, 0.3) -- ++(0, -0.3);
\end{tikzpicture}} \\
 \mathscr{D}(b_1)  = par^{p, q, r}_s (tensor^{p, q}_r (ax_p, ax_q)) & b_1  \equiv \raisebox{-0.7cm}{\begin{tikzpicture}
\node (tenb1) at (1,1) {$\bigotimes$};
\node (parb1) at (0,0.5) {$\parr$};
\node (ax1b1) at (0.2,1.5) {$\bullet$};
\node (ax2b1) at (0.7,1.5) {$\bullet$};
\draw (0.2,1.5) to [bend left=20] (tenb1);
\draw (0.2,1.5) edge[out=180, in=170] (parb1);
\draw (0.7,1.5) .. controls(0.5,1.5).. (parb1);
\draw (0.7,1.5) .. controls(0.9,1.5) .. (tenb1);
\draw (tenb1) to [bend left] (parb1);
\draw (0, 0.3) -- ++(0, -0.3);
\end{tikzpicture}} \\
\end{eqnarray*}
\par We write $\overrightarrow{b}$ for $b_{i_1}, \hdots, b_{i_n}$ for $i \in \{0, 1\}$.
\end{defi}
\par As we can see, $b_0$ and $b_1$ differ on their planarity: descriptions and proof nets exhibit the exchanges that were kept implicit in derivations.
\begin{defi}[Boolean proof nets \cite{terui04}] \label{bool_func}
A \textit{Boolean proof net with $n$ inputs} is a proof net $P(\overrightarrow{p})$ of type
\[\vdash p_1 : \textbf{B}^{\bot}[A_1], \hdots, p_n : \textbf{B}^{\bot}[A_n], s : \otimes^{1+m} ( \textbf{B}[A], D_1, \hdots, D_m)\]
\par Given $\overrightarrow{b}$ of length $n$, $P( \overrightarrow{b})$ is obtained by connecting with cuts $p_j$ to $b_{i_j}$ for all $1 \leq j \leq n$.
\par $P( \overrightarrow{b}) \rightrightarrows^* Q$ where $Q$ is unique, cut-free and for some descriptions $Q_1, \hdots, Q_n$ described by \[tensor(\mathscr{D}(b_i), Q_1, \hdots, Q_m)\text{ for }i \in \{0, 1\}.\] We write $P(\overrightarrow{b}) \to_{ev.} b_i$.
\par $P( \overrightarrow{p})$ \textit{represents a Boolean function} $f^n$ if for all $w \equiv i_1 \hdots i_n \in \{0, 1\}^n$, $P(b_{i_1}, \hdots, b_{i_n}) \to_{ev.}b_{f(w)}$.
\par We may easily define \textit{families of Boolean proof nets} and \textit{language accepted by a family of Boolean proof nets}.
\par The tensor indexed with $s$ in the type is the \textit{result tensor}: it collects the result of the computation on its first auxiliary port and the \textit{garbage} -- here named $D_1, \hdots, D_m$ -- on its other auxiliary ports.
\end{defi}
\begin{defi}[Direct Connection Language for proof nets \cite{mogbil-uniform}] \label{LDC_pn}
Given $P = (P_n )_{n \in \mathbb{N}}$ a family of Boolean proof nets, its \textit{Direct Connection Language} --~written $L_{DC}(P)$~-- is the set of tuples  $< y, \overline{g}, \overline{p}, \overline{b}>$ where for $|y| = n$: $g$ is a link in $P_n$, of sort $b$ if $p = \epsilon$ else the $p^{th}$ premise of $g$ is the link $b$.
\par If $<y, \overline{p}, 0, \overline{b}>$ or $<y, \overline{b}, 0, \overline{p}>$ belong to $L_{DC}(P)$, there is a cut between $b$ and $p$ in $C_{|y|}$.
\end{defi}
\section{Proof circuits}\label{sec5}
\par A proof circuit is a Boolean proof net (\autoref{pcbpn}) made out of pieces (\autoref{def_pieces}) which represents Boolean functions, constants or duplicates values. Garbage is manipulated in an innovative way, \autoref{exes} should help to understand the mechanism of computation.
\par From now on every edge represented by
\raisebox{-0.2cm}{
\begin{tikzpicture}
\draw[{-[}] (0,0) -- (0.5,0.5);
\end{tikzpicture}
}
is connected on its right to an auxiliary port numbered with an integer other than $1$ of the result tensor: it carries a piece of garbage.
\begin{defi}[Pieces]\label{def_pieces}
We present in the \autoref{pieces} the set of \textit{pieces} at our disposal. Entries are labeled with $e$, exits with $s$ and garbage with $g$. Edges labeled $b_k$ -- for $k \in \{0, 1\}$ -- are connected to the edge labeled $s$ of the piece $b_k$.
\vfill
\setlength{\LTleft}{-5cm plus 1 fill}
\setlength{\LTright}{-5cm plus 1 fill}
\begin{longtable}{|m{0.1\textwidth}m{0.4\textwidth}m{0.1\textwidth}m{0.4\textwidth}|}
\caption{Pieces} \label{pieces}
\endfirsthead
\multicolumn{4}{c}
{{ Table 1: Pieces -- continued from previous page}} \\
\endhead
\multicolumn{4}{l}{We set $2 \leqslant j \leqslant i$. }  \\
\hline
$b_0 \equiv$ &
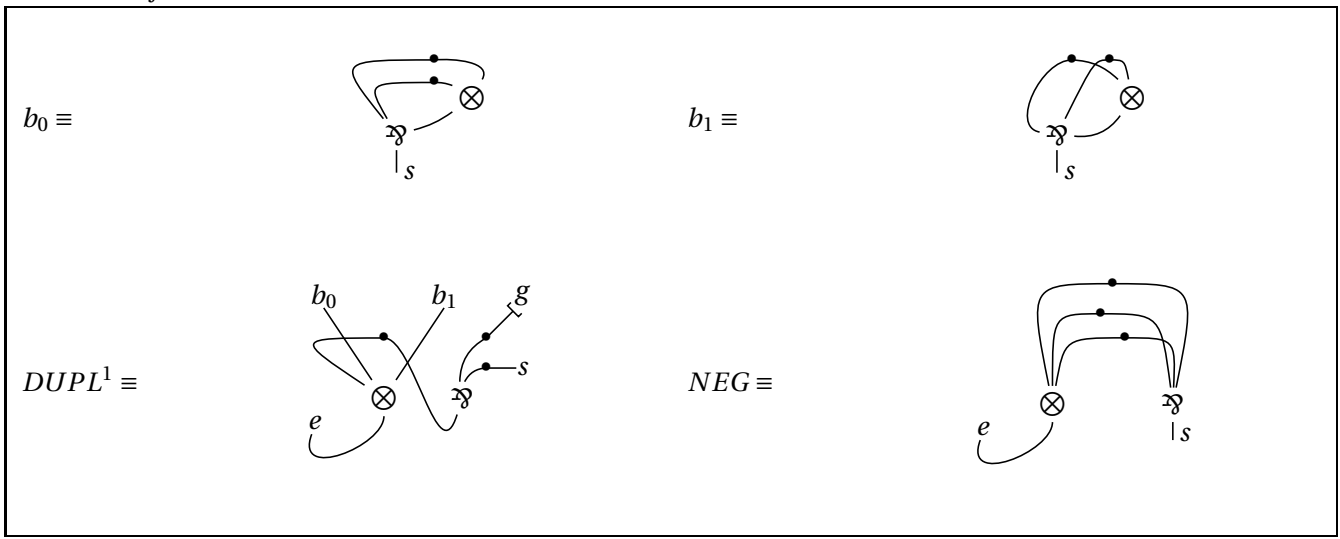
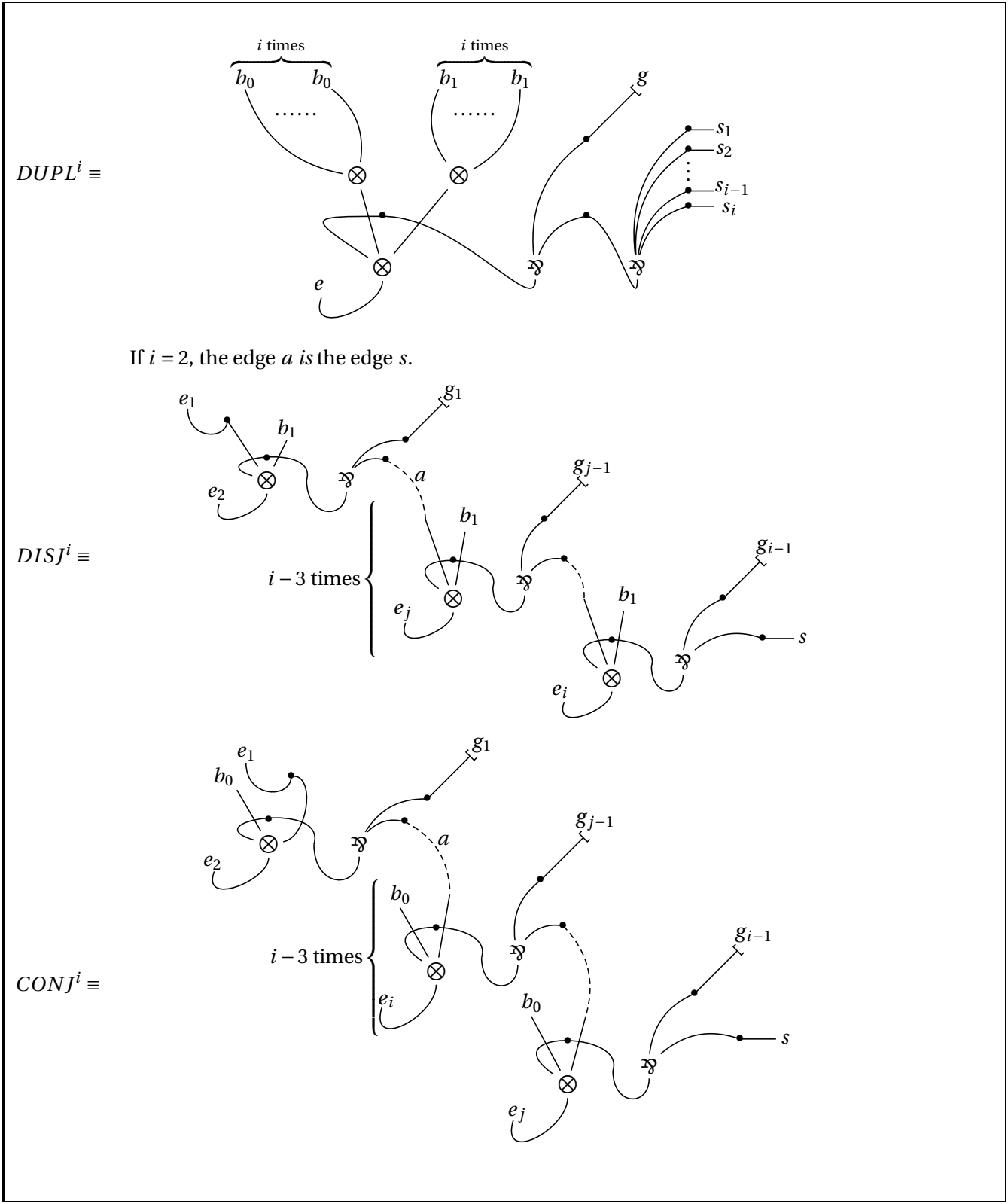
\begin{center}
\begin{tikzpicture}
\node (tenb0) at (1,1) {$\bigotimes$};
\node (parb0) at (0,0.5) {$\parr$};
\node (ax1b0) at (0.5,1.2) {$\bullet$};
\node (ax2b0) at (0.5,1.5) {$\bullet$};
\draw (0.5,1.5)  edge[out=5, in=60] (tenb0);
\draw (0.5,1.5) .. controls (-0.8,1.5) .. (parb0);
\draw (0.5,1.2) .. controls (-0.4,1.2) .. (parb0);
\draw (0.5,1.2) to [bend left=10] (tenb0);
\draw (tenb0) to [bend left=10] (parb0);
\draw (0, 0.3) node[pos=0.05, right]{$s$} -- ++(0, -0.3);
\end{tikzpicture}
\end{center}
&
$b_1 \equiv$ &
\begin{center}
\begin{tikzpicture}
\node (tenb1) at (1,1) {$\bigotimes$};
\node (parb1) at (0,0.5) {$\parr$};
\node (ax1b1) at (0.2,1.5) {$\bullet$};
\node (ax2b1) at (0.7,1.5) {$\bullet$};
\draw (0.2,1.5) to [bend left=20] (tenb1);
\draw (0.2,1.5) edge[out=180, in=170] (parb1);
\draw (0.7,1.5) .. controls(0.5,1.5).. (parb1);
\draw (0.7,1.5) .. controls(0.9,1.5) .. (tenb1);
\draw (tenb1) to [bend left] (parb1);
\draw (0, 0.3) node[pos=0.05, right]{$s$}  -- ++(0, -0.3);
\end{tikzpicture}
\end{center}  \\
$DUPL^1 \equiv$ &
\begin{center}
\begin{tikzpicture}[scale=0.8]
\node (ten) at (1.5,0.5) {$\bigotimes$};
\node (ax) at (1.5, 1.5) {$\bullet$};
\node (b0) at (0.5, 2.2) {$b_0$};
\node (b1) at (2.5, 2.2) {$b_1$};
\draw (0.5, 2) to (ten);
\draw (2.5, 2) to (ten);
\draw (1.5,1.5) .. controls (0,1.5).. (ten);
\node (par) at (2.8, 0.5) {$\parr$};
\draw (1.5, 1.5) .. controls (2, 1.5) and (2.4, -0.8) .. (par);
\node (ax2) at (3.2, 1.5) {$\bullet$};
\node (ax3) at (3.2, 1) {$\bullet$};
\draw [{-[}] (3.2, 1.5) to (3.7, 2);
\draw (3.2, 1) to (3.7, 1);
\draw (par) to [bend left] (3.2, 1.5);
\draw (par) to [bend left] (3.2, 1);
\node (s) at (3.8, 1) {$s$};
\draw (ten) .. controls ++(0, -0.8) and ++(-0.3, -0.8) .. node[above=12pt, left=5pt]{$e$} ++(-1.2, -0.6);
\node (g) at (3.8, 2.2) {$g$};
\end{tikzpicture}
\end{center}  &
$NEG \equiv$ &
\begin{center}
\begin{tikzpicture}[scale=0.8]
\node (ax1) at (1,2) {$\bullet$};
\node (ax2) at (0.8,1.5) {$\bullet$};
\node (ax3) at (1.2,1.1) {$\bullet$};
\node (ten) at (0,0) {$\bigotimes$};
\node (par) at (2,0) {$\parr$};
\draw (1,2) [-] ..controls ++(-1.4, 0).. (ten);
\draw (0.8,1.5)  [-] ..controls ++(-0.8, 0).. (ten);
\draw (1.2,1.1)  [-] ..controls ++(-1, 0).. (ten);
\draw (1,2)  [-] ..controls ++(1.4, 0).. (par);
\draw (0.8,1.5)  [-] ..controls ++(1, 0).. (par);
\draw (1.2,1.1)  [-] ..controls ++(0.85, 0).. (par);
\draw (2, -0.3) -- ++(0, -0.3);
\draw (ten) .. controls ++(0, -0.8) and ++(-0.3, -0.8) .. node[above=12pt, left=5pt]{$e$} ++(-1.2, -0.6);
\node (e) at (2.2, -0.5) {$s$};
\end{tikzpicture}
\end{center}
\\
\hline
$DUPL^i \equiv$ &
\multicolumn{3}{m{0.8\textwidth}|}{
\begin{center}
\begin{tikzpicture}[scale=0.9]
\node (ten) at (1.5,0.5) {$\bigotimes$};
\node (ax) at (1.5, 1.5) {$\bullet$};
\node (ten1) at (1, 2.3) {$\bigotimes$};
\node (ten2) at (3, 2.3) {$\bigotimes$};
\draw (ten1) to (ten);
\draw (ten2) to (ten);
\draw (1.5,1.5) .. controls (0,1.5).. (ten);
\draw (ten) .. controls ++(0, -0.8) and ++(-0.3, -0.8) .. node[above=15pt, left=9pt]{$e$} ++(-1.2, -0.6);
\node (b0) at (-1.2, 4.2) {$b_0$};
\node (b0) at (0.3, 4.2) {$b_0$};
\draw (ten1) to [bend left] (-1.2, 4);
\draw (ten1) to [bend right] (0.5, 4);
\node (acc) at (-0.5, 4.7) {$\overbrace{~~~~~~~~~~~~~~~~~~~~~~}^{i\text{ times}}$};
\node (b1) at (2.8, 4.2) {$b_1$};
\node (b1) at (4.2, 4.2) {$b_1$};
\draw (ten2) to [bend left] (2.6, 4);
\draw (ten2) to [bend right] (4.2, 4);
\node (acc) at (3.5, 4.7) {$\overbrace{~~~~~~~~~~~~~~~~~~~~~}^{i\text{ times}}$};
\node (dots) at (-0.2, 3.5) {$\hdots \hdots$};
\node (dots) at (3.3, 3.5) {$\hdots \hdots$};
\node (par) at (4.5,0.5) {$\parr$};
\draw (1.5, 1.5) .. controls (3, 1.5) and (4.5, -0.5) .. (par);
\node (ax) at (5.5, 1.5) {$\bullet$};
\node (par2) at (6.5,0.5) {$\parr$};
\draw (5.5, 1.5) .. controls (6, 1.5) and (6.5, -0.5) .. (par2);
\draw (5.5, 1.5) to [bend right] (par);
\draw (par) to [bend left] (5.5, 3);
\node (ax) at (5.5, 3) {$\bullet$};
\draw [{-[}] (5.5, 3) to  (6.5, 4);
\node (g) at (6.6, 4.2) {$g$};
\draw (par2) to [bend left] (7.5, 1.7);
\draw (par2) to [bend left] (7.5, 2);
\draw (par2) to [bend left] (7.5, 2.8);
\draw (par2) to [bend left] (7.5, 3.2);
\node (dots) at (7.5, 2.5) {$\vdots$};
\node (ax) at (7.5, 1.7) {$\bullet$};
\node (ax) at (7.5, 2) {$\bullet$};
\node (ax) at (7.5, 2.8) {$\bullet$};
\node (ax) at (7.5, 3.2) {$\bullet$};
\draw (7.5, 1.7) -- ++(0.5, 0);
\draw (7.5, 2) -- ++(0.5, 0);
\draw (7.5, 3.2) -- ++(0.5, 0);
\draw (7.5, 2.8) -- ++(0.5, 0);
\node (s1) at (8.2, 3.2) {$s_1$};
 \node (s1) at (8.2, 2.8) {$s_{2}$};
\node (s1) at (8.3, 1.65) {$s_{i}$};
\node (s1) at (8.35, 2.05) {$s_{i-1}$};
\end{tikzpicture}
\vspace{-1.5em}
\end{center}}
\\
$DISJ^i \equiv$ &
\multicolumn{3}{m{0.8\textwidth}|}{
If $i=2$, the edge $a$ \textit{is} the edge $s$.
\begin{center}
\begin{tikzpicture}[scale=0.7]
\node (par) at (-0.5, 1) {$\parr$};
\node (ax1) at (0.5, 1.5) {$\bullet$};
\draw (par) to [bend left] (0.5, 1.5);
\node (ax2) at (1, 2) {$\bullet$};
\draw (par) to [bend left] (1,2);
\draw [{-[}] (1,2) to  (2,3);
\node (g1) at (2.2, 3.2) {$g_1$};
\draw (par) .. controls (-0.5, 0) and (-1.5, 0) ..  (-1.5, 1);
\node (b1) at (-2, 2.3) {$b_1$};
\node (ten) at (-2.5, 1) {$\bigotimes$};
\draw (-2,2) to (ten);
\node (ax) at (-2.5, 1.54) {$\bullet$};
\draw (ten) to (-3.5, 2.5);
\draw (ten) .. controls (-4.2, 1.5) and (-1.2, 2) .. (-1.5, 1);
\node (e1) at (-4.5,3) {$e_1$};
\node (ax) at (-3.5, 2.5) {$\bullet$};
\draw (-4.5, 2.8) .. controls (-4.5, 2) and (-3.5, 2) ..(-3.5, 2.5);
\draw (ten) .. controls ++(0, -0.8) and ++(-0.3, -0.8) .. node[above=12pt, left=4pt]{$e_2$} ++(-1.2, -0.6);
\node (e2) at (1.35,1.1) {$a$};
\draw[densely dashed] (0.5, 1.5) to [bend left] (1.5, 0);
\node (teni) at (2.2, -2) {$\bigotimes$};
\draw (1.5,0) to (teni);
\draw (teni) to (2.5, -0.3);
\node (b1) at (2.6, 0.1) {$b_1$};
\node (ax) at (2.2, -1.05) {$\bullet$};
\draw (teni) .. controls (0.5,-0.8) and (3.5, -0.8) .. (3.2, -1.7);
\draw (teni) .. controls ++(0, -0.8) and ++(-0.3, -0.8) .. node[above=12pt, left=4pt]{$e_j$} ++(-1.2, -0.6);
\node (par2) at (4,-1.6) {$\parr$};
\draw (par2) .. controls (4, -2.5) and (3.2, -2.5) ..  (3.2, -1.7);
\draw (par2) to [bend left] (5, -1);
\node (ax4) at (4.5, 0) {$\bullet$};
\draw (par2) to [bend left] (4.5, 0);
\draw [{-[}] (4.5,0) to (5.5, 1);
\node (g2) at (5.7, 1.3) {$g_{j-1}$};
\draw [densely dashed] (5, -1) to [bend left] (5.5, -2);
\node (ax3) at (5, -1) {$\bullet$};
\begin{scope}[xshift=4cm, yshift=-2cm]
\node (teni2) at (2.2, -2) {$\bigotimes$};
\draw (1.5,0) to (teni2);
\draw (teni2) to (2.5, -0.3);
\node (b1) at (2.6, 0.1) {$b_1$};
\node (ax) at (2.2, -1.05) {$\bullet$};
\draw (teni2) .. controls (0.5,-0.8) and (3.5, -0.8) .. (3.2, -1.7);
\draw (teni2) .. controls ++(0, -0.8) and ++(-0.3, -0.8) .. node[above=12pt, left=5pt]{$e_i$} ++(-1.2, -0.6);
\node (par2) at (4,-1.6) {$\parr$};
\draw (par2) .. controls (4, -2.5) and (3.2, -2.5) ..  (3.2, -1.7);
\node (ax3) at (6, -1) {$\bullet$};
\draw (par2) to [bend left] (6, -1);
\node (ax4) at (5, 0) {$\bullet$};
\draw (par2) to [bend left] (5, 0);
\draw [{-[}] (5,0) to (6, 1);
\node (g3) at (6.3, 1.3) {$g_{i-1}$};
\draw (6, -1) -- ++(0.8, 0);
\node (s) at (7, -1) {$s$};
\end{scope}
\node (ac) at (-0.8, -1.5) {$
\mbox{$i-3$ times}
\left\{ \begin{array}{l}
\,\\
\,\\
\,\\
\,\\
\,\\
\,\\
\end{array}\right.
$};
\end{tikzpicture}
\end{center}
}
\\
$CONJ^i \equiv$ &
\multicolumn{3}{m{0.8\textwidth}|}{
\begin{center}
\vspace{-3.5em}
\begin{tikzpicture}[scale=0.8]
\node (ten) at (-2.5, 1) {$\bigotimes$};
\node (par) at (-0.5, 1) {$\parr$};
\node (e1) at (-3,3) {$e_1$};
\draw (-3, 2.8) .. controls (-3, 2) and (-2, 2) ..(-2, 2.5);
\node (ax) at (-2, 2.5) {$\bullet$};
\draw (-2,2.5) edge[out=10, in=10] (ten);
\node (ax) at (-3.5, 2.5) {$b_0$};
\draw (ten) to (-3.2, 2.2);
\draw (par) to [bend left] (0.5, 1.5);
\node (ax2) at (1, 2) {$\bullet$};
\draw (par) to [bend left] (1,2);
\draw [{-[}] (1,2) to  (2,3);
\node (g1) at (2.2, 3.2) {$g_1$};
\draw (par) .. controls (-0.5, 0) and (-1.5, 0) ..  (-1.5, 1);
\node (ax1) at (0.5, 1.5) {$\bullet$};
\node (ax) at (-2.5, 1.54) {$\bullet$};
\draw (ten) .. controls ++(0, -0.8) and ++(-0.3, -0.8) .. node[above=12pt, left=5pt]{$e_2$} ++(-1.2, -0.6);
\node (e2) at (1.35,1.1) {$a$};
\draw (ten) .. controls (-4.2, 1.5) and (-1.2, 2) .. (-1.5, 1);
\draw[densely dashed] (0.5, 1.5) to [bend left] (1.5, 0);
\begin{scope}[xshift=-1cm, yshift=0.2cm]
\node (teni) at (2.2, -2) {$\bigotimes$};
\draw (1.4,-0.6) to (teni);
\node (ax) at (1.4, -0.3) {$b_0$};
\draw (teni) to (2.5, -0.3);
\node (ax) at (2.2, -1.05) {$\bullet$};
\draw (teni) .. controls (0.5,-0.8) and (3.5, -0.8) .. (3.2, -1.7);
\node (par2) at (4,-1.6) {$\parr$};
\draw (par2) .. controls (4, -2.5) and (3.2, -2.5) ..  (3.2, -1.7);
\draw (par2) to [bend left] (5, -1);
\node (ax4) at (4.5, 0) {$\bullet$};
\draw (par2) to [bend left] (4.5, 0);
\draw [{-[}] (4.5,0) to (5.5, 1);
\node (g2) at (5.7, 1.3) {$g_{j-1}$};
\draw [densely dashed] (5, -1) to [bend left] (5.5, -3);
\node (ax3) at (5, -1) {$\bullet$};
\end{scope}
\begin{scope}[xshift=1.9cm, yshift=-2.3cm]
\node (teni2) at (2.2, -2) {$\bigotimes$};
\draw (1.4,-0.5) to (teni2);
\draw (teni2) to (2.6, -0.5);
\node (b0) at (1.4, -0.2) {$b_0$};
\node (ax) at (2.2, -1.05) {$\bullet$};
\draw (teni2) .. controls (0.5,-0.8) and (3.5, -0.8) .. (3.2, -1.7);
\draw (teni2) .. controls ++(0, -1) and ++(-0.3, -1) .. node[above=12pt, left=2pt]{$e_j$} ++(-1.2, -0.8);
\draw (teni) .. controls ++(0, -1) and ++(-0.3, -1) .. node[above=12pt, left=2pt]{$e_i$} ++(-1.2, -0.8);
\node (par2) at (4,-1.6) {$\parr$};
\draw (par2) .. controls (4, -2.5) and (3.2, -2.5) ..  (3.2, -1.7);
\node (ax3) at (6, -1) {$\bullet$};
\draw (par2) to [bend left] (6, -1);
\node (ax4) at (5, 0) {$\bullet$};
\draw (par2) to [bend left] (5, 0);
\draw [{-[}] (5,0) to (6, 1);
\node (g3) at (6.3, 1.3) {$g_{i-1}$};
\draw (6, -1) -- ++(0.8, 0);
\node (s) at (7, -1) {$s$};
\end{scope}
\node (ac) at (-1, -1.5) {$
\mbox{$i-3$ times}
\left\{ \begin{array}{l}
\,\\
\,\\
\,\\
\,\\
\,\\
\,\\
\end{array}\right.
$};
\end{tikzpicture}
\end{center}
} \\
\hline
\end{longtable}

\par A \textit{piece $\mathscr{P}$ with $i \geqslant 0$ entries, $j \geqslant 1$ exits and $k \geqslant 0$ garbage} is one of the piece in the \autoref{pieces}, where $i$ edges are labeled with $e_1, \hdots, e_i$, $j$ edges are labeled with $s_1, \hdots, s_j$ and $k$ edges go to the result tensor.
\par We have $\mathscr{P} \in \{b_0, b_1, NEG, \{DUPL^i\}_{i \geqslant 1}, \{DISJ^i\}_{i \geqslant 2}, \{CONJ^i\}_{i \geqslant 2} \}$.
\par To \textit{compose two pieces} $\mathscr{P}_1$ and $\mathscr{P}_2$ we connect an exit of $\mathscr{P}_1$ to an entry of $\mathscr{P}_2$. It is not allowed to loop: we can not connect an entry and an exit belonging to the same piece.
\par An entry (resp. an exit) that is not connected to an exit (resp. an entry) of another piece is said to be \textit{unconnected}.
\end{defi}
\begin{defi}[Proof circuits]
\par A \textit{proof circuit $\mathscr{C}_n(\overrightarrow{p})$ with $n$ inputs and one output} is obtained by composing pieces such that $n$ entries and one exit are unconnected. If no garbage is created we add a $DUPL^1$-piece connected to the unconnected exit to produce some artificially. Then we add a result tensor whose first edge is connected to the exit -- which is also the output of the proof circuit -- and the others to the garbage. We then label every unconnected entries with $p_1, \hdots, p_n$: those are the inputs of the proof circuit.
\par Given $\overrightarrow{b}$ of length $n$, $\mathscr{C}_n( \overrightarrow{b})$ is obtained by connecting with cuts $p_j$ to $b_{i_j}$ for all $1 \leq j \leq n$.
\end{defi}
\begin{fac}\label{pcbpn}
 Every proof circuit is a Boolean proof net.
\end{fac}
\begin{proo}
 We prove this fact with a contractibility criterion \cite{danos90}, by induction on the height of the pieces of the proof circuit (counted as the number of pieces from the considered piece to the result tensor). As a proof circuit can always be typed with
\[\vdash \textbf{B}^{\bot}[A_1], \hdots, \textbf{B}^{\bot}[A_n], \otimes^{1+m} ( \textbf{B}[A], D_1, \hdots, D_m)\]
it is a Boolean proof net.
\end{proo}
\par This fact establishes that proof circuits normalize and output a value, and that it is possible to represent Boolean functions with them.
\par \textit{Families of proof circuits} and \textit{acceptation of a language by a family of proof circuits} are defined as usual.
\begin{exes}\label{exes}
\par We present briefly two examples of computation: the normalization of a piece $b_0$ connected to a $NEG$-piece (\autoref{ex1}, \autopageref{ex1}) and how the conditional works (\autoref{ex2}, \autopageref{ex2}). Conditional is the core of the computation, we find this pattern in every piece except $NEG$ and the constants.
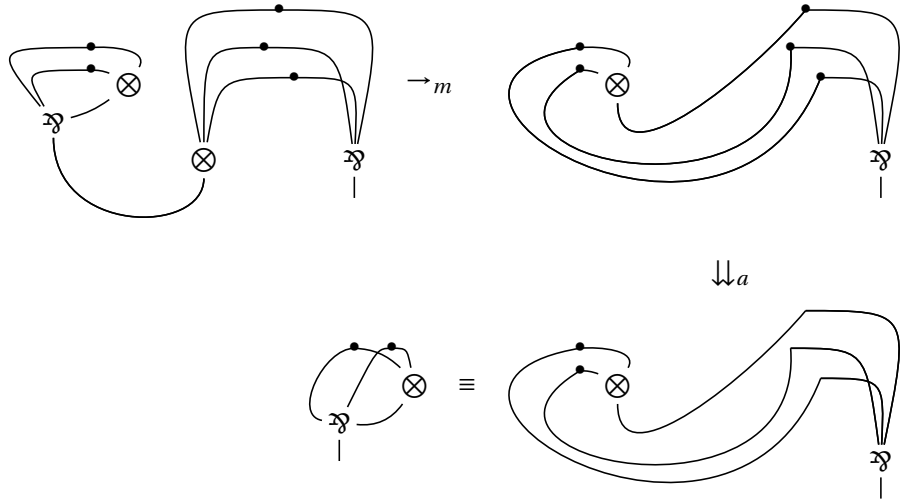
\begin{figure}[ht]
\begin{center}
\begin{tikzpicture}
\node (tenb0) at (1,1) {$\bigotimes$};
\node (parb0) at (0,0.5) {$\parr$};
\node (ax1b0) at (0.5,1.2) {$\bullet$};
\node (ax2b0) at (0.5,1.5) {$\bullet$};
\draw (0.5,1.5)  edge[out=5, in=60] (tenb0);
\draw (0.5,1.5) .. controls (-0.8,1.5) .. (parb0);
\draw (0.5,1.2) .. controls (-0.4,1.2) .. (parb0);
\draw (0.5,1.2) to [bend left=10] (tenb0);
\draw (tenb0) to [bend left=10] (parb0);
\node (ax1) at (3,2) {$\bullet$};
\node (ax2) at (2.8,1.5) {$\bullet$};
\node (ax3) at (3.2,1.1) {$\bullet$};
\node (ten) at (2,0) {$\bigotimes$};
\node (par) at (4,0) {$\parr$};
\draw (3,2) [-] ..controls ++(-1.4, 0).. (ten);
\draw (2.8,1.5)  [-] ..controls ++(-0.8, 0).. (ten);
\draw (3.2,1.1)  [-] ..controls ++(-1, 0).. (ten);
\draw (3,2)  [-] ..controls ++(1.4, 0).. (par);
\draw (2.8,1.5)  [-] ..controls ++(1, 0).. (par);
\draw (3.2,1.1)  [-] ..controls ++(0.85, 0).. (par);
\draw (par) -- ++(0, -0.5);
\draw (0, 0.3) .. controls (0, -1) and (2, -1) .. (ten);
\draw (0, 0.3) .. controls (0, -1) and (2, -1) .. (ten);
\node (fleche1) at (5, 1) {$\to_m$};
\node (ax4) at (7, 1.2) {$\bullet$};
\node (ax5) at (7, 1.5) {$\bullet$};
\node (tenbis) at (7.5, 1) {$\bigotimes$};
\node (parbis) at (11, 0) {$\parr$};
\draw (7, 1.2) to [bend left=10] (tenbis);
\draw (7, 1.5) edge[out=5, in=60] (tenbis);
\node (ax6) at (10,2) {$\bullet$};
\node (ax7) at (9.8,1.5) {$\bullet$};
\node (ax8) at (10.2,1.1) {$\bullet$};
\draw (7, 1.2) .. controls (5, 0) and (10, -1) .. (9.8, 1.5);
\draw (7, 1.5) .. controls (4, 1) and (9, -2) .. (10.2, 1.1);
\draw (tenbis) edge[out=270, in=230] (10, 2);
\draw (7, 1.2) .. controls (5, 0) and (10, -1) .. (9.8, 1.5);
\draw (7, 1.5) .. controls (4, 1) and (9, -2) .. (10.2, 1.1);
\draw (tenbis) edge[out=270, in=230] (10, 2);
\draw (10,2)  ..controls ++(1.4, 0).. (parbis);
\draw (9.8,1.5) ..controls ++(1, 0).. (parbis);
\draw (10.2,1.1) ..controls ++(0.85, 0).. (parbis);
\draw (parbis) -- ++(0, -0.5);
\node (fleche2) at (9, -1.5) {$\downdownarrows_a$};
\begin{scope}[xshift = 0cm, yshift = -4cm]
\node (ax4) at (7, 1.2) {$\bullet$};
\node (ax5) at (7, 1.5) {$\bullet$};
\node (tenbis) at (7.5, 1) {$\bigotimes$};
\draw (7, 1.2) to [bend left=10] (tenbis);
\draw (7, 1.5) edge[out=5, in=60] (tenbis);
\draw (7, 1.2) .. controls (5, 0) and (10, -1) .. (9.8, 1.5);
\draw (7, 1.5) .. controls (4, 1) and (9, -2) .. (10.2, 1.1);
\draw (tenbis) edge[out=270, in=230] (10, 2);
\node (parbis) at (11, 0) {$\parr$};
\draw (10,2)  ..controls ++(1.4, 0).. (parbis);
\draw (9.8,1.5) ..controls ++(1, 0).. (parbis);
\draw (10.2,1.1) ..controls ++(0.85, 0).. (parbis);
\draw (parbis) -- ++(0, -0.5);
\draw (10,2) ..controls ++(1.4, 0).. (parbis);
\draw (9.8,1.5) ..controls ++(1, 0).. (parbis);
\draw (10.2,1.1) ..controls ++(0.85, 0).. (parbis);
\end{scope}
\node at (5.5, -3) {$\equiv$};
\begin{scope}[xshift = 3.8cm, yshift = -4cm]
\node (tenb1) at (1,1) {$\bigotimes$};
\node (parb1) at (0,0.5) {$\parr$};
\node (ax1b1) at (0.2,1.5) {$\bullet$};
\node (ax2b1) at (0.7,1.5) {$\bullet$};
\draw (0.2,1.5) to [bend left=20] (tenb1);
\draw (0.2,1.5) edge[out=180, in=170] (parb1);
\draw (0.7,1.5) .. controls(0.5,1.5).. (parb1);
\draw (0.7,1.5) .. controls(0.9,1.5) .. (tenb1);
\draw (tenb1) to [bend left] (parb1);
\draw (parb1) -- ++(0, -0.5);
\end{scope}
\end{tikzpicture}
\vspace{-5em}
\end{center}
\caption{$b_0$ connected to $NEG$ normalizes to $b_1$}\label{ex1}
\end{figure}
\begin{figure}[ht]
\begin{center}
\begin{tikzpicture}
\draw [dashed] (-0.5, 0) rectangle (1.2, 1.8);
\node (tenb1) at (1,1) {$\bigotimes$};
\node (parb1) at (0,0.5) {$\parr$};
\node (ax1b1) at (0.2,1.5) {$\bullet$};
\node (ax2b1) at (0.7,1.5) {$\bullet$};
\draw (0.2,1.5) to [bend left=20] (0.8, 1.2);
\draw (0.7,1.5) .. controls(0.9,1.5) .. (1, 1.25);
\draw (0.2,1.5) edge[out=180, in=170] (parb1);
\draw (0.7,1.5) .. controls(0.5,1.5).. (parb1);
\draw (tenb1) to [bend left] (parb1);
\node (tenseur) at (2, 0) {$\bigotimes$};
\draw (parb1) edge[out=270, in=270] (tenseur);
\node (axten) at (2, 1) {$\bullet$};
\draw (tenseur) edge[out=150, in=150] (2, 1);
\draw (2, 1) edge[out=0, in=90] (3, 0);
\begin{scope}[xshift=1cm, yshift=2cm]
\node (tenb0) at (1,1) {$\bigotimes$};
\node (parb0) at (0,0.5) {$\parr$};
\node (ax1b0) at (0.5,1.2) {$\bullet$};
\node (ax2b0) at (0.5,1.5) {$\bullet$};
\draw (0.5,1.5)  edge[out=5, in=60] (1.1, 1.25);
\draw (0.5,1.5) .. controls (-0.8,1.5) .. (parb0);
\draw (0.5,1.2) .. controls (-0.4,1.2) .. (parb0);
\draw (0.5,1.2) to [bend left=10] (0.75, 1.2);
\draw (tenb0) to [bend left=10] (parb0);
\end{scope}
\draw (tenseur) to (parb0);
\begin{scope}[xshift=3cm, yshift=2cm]
\node (tenb1') at (1,1) {$\bigotimes$};
\node (parb1') at (0,0.5) {$\parr$};
\node (ax1b1') at (0.2,1.5) {$\bullet$};
\node (ax2b1') at (0.7,1.5) {$\bullet$};
\draw (0.2,1.5) edge[out=180, in=170] (parb1');
\draw (0.7,1.5) .. controls(0.5,1.5).. (parb1');
\draw (0.2,1.5) to [bend left=20] (0.8, 1.2);
\draw (0.7,1.5) .. controls(0.9,1.5) .. (1, 1.25);
\draw (tenb1') to [bend left] (parb1');
\end{scope}
\draw (tenseur) to (parb1');
\node (fleche) at (4.5, 2) {$\to_m$};
\begin{scope}[xshift=5cm, yshift=0]
\begin{scope}[xshift=1cm, yshift=2cm]
\node (tenb0') at (1,1) {$\bigotimes$};
\node (parb0') at (0,0.5) {$\parr$};
\node (ax1b0') at (0.5,1.2) {$\bullet$};
\node (ax2b0') at (0.5,1.5) {$\bullet$};
\draw (0.5,1.5) edge[out=5, in=60] (1.1, 1.25);
\draw (0.5,1.2) to [bend left=10] (0.75, 1.2);
\draw (0.5,1.5) .. controls (-0.8,1.5) .. (parb0');
\draw (0.5,1.2) .. controls (-0.4,1.2) .. (parb0');
\draw (tenb0') to [bend left=10] (parb0');
\end{scope}
\begin{scope}[xshift=3cm, yshift=2cm]
\node (tenb1) at (1,1) {$\bigotimes$};
\node (parb1) at (0,0.5) {$\parr$};
\node (ax1b1) at (0.2,1.5) {$\bullet$};
\node (ax2b1) at (0.7,1.5) {$\bullet$};
\draw (0.2,1.5) edge[out=180, in=170] (parb1);
\draw (0.7,1.5) .. controls(0.5,1.5).. (parb1);
\draw (0.2,1.5) to [bend left=20] (0.8, 1.2);
\draw (0.7,1.5) .. controls(0.9,1.5) .. (1, 1.25);
\draw (tenb1) to [bend left] (parb1);
\end{scope}
\node (tenb0) at (1,1) {$\bigotimes$};
\node (ax1b0) at (0.2,1.5) {$\bullet$};
\node (ax2b0) at (0.7,1.5) {$\bullet$};
\draw (0.2,1.5) to [bend left=20] (0.8, 1.2);
\draw (0.7,1.5) .. controls(0.9,1.5) .. (1, 1.25);
\node (axten) at (2, 1) {$\bullet$};
\draw (2, 1) edge[out=0, in=90] (3, 0);
\draw (tenb0) edge[out=270, in=90] (2, 1);
\draw (0.2, 1.5) .. controls (-1.5, -0.5) and (3, -1) .. (parb1);
\draw (0.7, 1.5) .. controls (-0.5, 0) and (3, -1) .. (parb0');
\end{scope}
\node (fleche2) at (9.5, 2) {$\rightrightarrows_a$};
\node (tenfinal) at (11.5, 1) {$\bigotimes$};
\begin{scope}[xshift=12.5cm, yshift=2cm]
\node (tenb0) at (1,1) {$\bigotimes$};
\node (parb0) at (0,0.5) {$\parr$};
\node (ax1b0) at (0.5,1.2) {$\bullet$};
\node (ax2b0) at (0.5,1.5) {$\bullet$};
\draw (0.5,1.5) .. controls (-0.8,1.5) .. (parb0);
\draw (0.5,1.2) .. controls (-0.4,1.2) .. (parb0);
\draw (0.5,1.5) edge[out=5, in=60] (1.1, 1.25);
\draw (0.5,1.2) to [bend left=10] (0.75, 1.2);
\draw (tenb0) to [bend left=10] (parb0);
\end{scope}
\begin{scope}[xshift=10.5cm, yshift=2cm]
\node (tenb1) at (1,1) {$\bigotimes$};
\node (parb1) at (0,0.5) {$\parr$};
\node (ax1b1) at (0.2,1.5) {$\bullet$};
\node (ax2b1) at (0.7,1.5) {$\bullet$};
\draw (0.2,1.5) edge[out=180, in=170] (parb1);
\draw (0.7,1.5) .. controls(0.5,1.5).. (parb1);
\draw (0.2,1.5) to [bend left=20] (0.8, 1.2);
\draw (0.7,1.5) .. controls(0.9,1.5) .. (1, 1.25);
\draw (tenb1) to [bend left] (parb1);
\end{scope}
\draw (tenfinal) to (parb1);
\draw (tenfinal) to (parb0);
\draw (tenfinal) -- ++(0, -1);
\end{tikzpicture} 
\end{center}
\par \small{The input (here in a dashed rectangle) is proof net of type $\textbf{B}$, and it “selects” -- according to its planarity or non-planarity-- during the normalization which one of $b_0$ or $b_1$ is connected to the first auxiliary port of the tensor and so is considered as the result -- the other being treated as garbage.}
\caption{The conditional, the core of the computation}\label{ex2}
\end{figure}
\end{exes}
\begin{rem}
 \par To \textit{compose two proof circuits} $\mathscr{C}_1$ and $\mathscr{C}_2$, we remove the result tensor of $\mathscr{C}_1$, identify the unconnected exit of $\mathscr{C}_1$ with the selected input of $\mathscr{C}_2$, and recollect all the garbage with the result tensor of $\mathscr{C}_2$. We then label the unconnected entries anew and obtain a proof circuit.
\end{rem}
\begin{defi}[$PCC^i$ (resp. $mBN^i$, \cite{mogbil-uniform})]
A language $X \subseteq \{0, 1\}^*$ belongs to the class $PCC^i$ (resp. $mBN^i$) if $X$ is accepted by a poly\-no\-mial-size, $\log^i$-depth uniform family of proof circuits (resp. of Boolean proof nets).
\par If we add "\textit{$UstConn_2$-pieces}" to the set of pieces, we may easily define $PCC^i(UstConn_2)$ and remark that for all $i \in \mathbb{N}$, $PCC^i \subseteq PCC^i(UstConn_2)$. \cite{terui04} proves that there exists Boolean proof nets of constant-depth and polynomial size that represent $UstConn_2$, so we do not define $mBN^i(\textit{UstConn}_2)$ because it would be easy to prove that this class is equal to $mBN^i$ for all $i \in \mathbb{N}$.
\end{defi}
\begin{rem}\label{mBNPCC}
 By \autoref{pcbpn} we have trivially that for all $i \in \mathbb{N}$, $PCC^i \subseteq mBN^i$.
\end{rem}
\begin{lem}\label{lem_prof}
 For all proof circuit $\mathscr{C}_n(\overrightarrow{p})$ and all $\overrightarrow{b}$, the cuts at maximum depth in $\mathscr{C}_n(\overrightarrow{b})$ are between the entry of a piece and a value (a constant $b_0$ or $b_1$, or an input $b_{i_j}$ for some $1 \leq j \leq n$).
\end{lem}
\begin{proo}
For every piece $\mathscr{P}$ of $\mathscr{C}_n(\overrightarrow{b})$ any cut connecting an entry is always of depth superior or equal to the maximal depth of the cuts connecting the exits. The cuts of $\mathscr{P}$ that do not connect an entry or an exit of a piece are always of depth inferior or equal to cuts connecting the entries.
\par The depths of the cut formulae slowly increase from the exit to the entry, and as the entries that are not connected to other pieces are connected to values, this lemma is proved.
\end{proo}
\section{Results}\label{sec6}
\par By using our proof circuits we prove anew the inclusions between $AC^i$ and logical classes of complexity and extend this inclusion to sublogarithmic classes of complexity.
\begin{defi}[Problem: Translation from $AC^i$ to $PCC^i$] \ 
\begin{center}
\begin{tabular}{l p{10cm}}
Input:&  $L_{DC}(C)$ for $C$ a family of Boolean circuits in $AC^i$.\\
Output:& $L_{DC}(\mathscr{C})$ for $\mathscr{C}$ a family of proof circuits in $PCC^i$, such that for all $n \in \mathbb{N}$, for all $\overrightarrow{b} \equiv b_{i_1}, \hdots, b_{i_n}$, $\mathscr{C}_n(\overrightarrow{b}) \to_{ev.} b_j$ iff $C_n(i_1, \hdots, i_n)$ evaluates to $j$.
\end{tabular}
\end{center}
\end{defi}
\begin{thm}\label{NCtoBC}
For all $i \in \mathbb{N}$, translation from $AC^i$ to $PCC^i$ belongs to $AC^0$.
\end{thm}
\begin{proo}
The translation from $C$ to $\mathscr{C}$ is obvious, it relies on coding: for every $n$, a first constant-depth circuit associate to every gate of $C_n$ the corresponding piece simulating its Boolean function. If the fan-out of this gate is $k>1$, a $DUPL^k$-piece is associated to the exit of the piece, and the pieces are connected as the gates. The input nodes are associated to the inputs of $\mathscr{C}_n$. A second constant-depth circuit recollects the only free exit and the garbage of the pieces and connects them to the result tensor. The composition of these two Boolean circuits produces a constant-depth Boolean circuit that builds proof circuits.
\par It is easy to check that $CONJ^k$, $DISJ^k$ and $NEG$ represent $\wedge^k$, $\vee^k$ and $\neg$ respectively. $DUPL^k$ duplicates a value $k$ times, $b_0$ and $b_1$ represent $0$ and $1$ by convention. The composition of these pieces does not raise any trouble: $\mathscr{C}_n$ effectively simulates $C_n$ on every input of size $n$.
\par Concerning the bounds: the longest path between an entry or a constant and the result tensor goes through at most $2 \times d(C_n)$ pieces and we know by \autoref{lem_prof} that the increase of the depth is linear in the number of pieces crossed. We conclude that $d(\mathscr{C}_n) \leqslant 2 \times 3 \times d(C_n)$ and that $\mathscr{C}_n$ normalizes in $O(d(C_n))$ parallel steps.
\par Concerning the size, by counting we know that a gate of fan-in $n$ and fan-out $m$ is simulated by a piece made of $O(m+n)$ links. As the number of edges in $C_n$ is bounded by $|C_n|^2$, the size of $\mathscr{C}_n$ is at most $O(|C_n|^2)$.
\end{proo}
\par A Boolean circuit with unbounded (resp. bounded) arity of size $s$ is translated by a Proof circuit of size quadratic (resp. linear) in $s$, whereas \cite{terui04} considers only unbounded Boolean circuits and translate them with Boolean proof nets of size $O(s^5)$. Our translation --~thanks mostly to the easier garbage collection~-- needs less computational power, is more clear and besides lower the size of the Boolean proof nets obtained.
\par Of course, we could naturally extend this translation to a translation from $AC^i(\textit{UstConn}_2)$ to $PCC^i(\textit{UstConn}_2)$ and still remain in $AC^0$. But it is of little interest to look for a sublogarithmic translation toward a class of complexity which is probably not strictly included in $L$.
\begin{fac}
As the reduction from $C$ to $\mathscr{C}$ is in $AC^0$, we know that this reduction is correct for Boolean circuit families in $AC^0$ and that every $\mathscr{C}$ obtained by this translation is uniform.
\end{fac}
\par This result brings a novelty in the study of the proof nets as a class of complexity, making them able to simulate very small classes of complexity born from the Boolean circuits.
\begin{thm}[Simulation]\label{mbntoPN}
For all $i \in \mathbb{N}$, for all Boolean proof net family $P = (P_n)_{n \in \mathbb{N}}$ in $mBN^i$, there exists a family of Boolean circuits $C = (C_n)_{n \in \mathbb{N}}$ in $AC^i(\textit{UstConn}_2)$ and a constant-depth circuit in $AC^0$ that given $L_{DC}(P)$ outputs $L_{DC}(C)$ such that for all $\overrightarrow{b} \equiv b_{i_1} \hdots b_{i_n}$, $P_n(\overrightarrow{b}) \to_{ev.} b_j$ iff $C_n(i_1, \hdots, i_n)$ evaluates to $j$.
\end{thm}
\begin{proo}
We know thanks to \cite{terui04} that for $r \in \{a, m, t\}$ an unbounded fan-in constant-depth circuit with $O(|P_n|^3)$ gates -- with $\textit{UstConn}_2$ gates to identify chains of axioms if $r = t$-- is able to reduce all the $r$-cuts of $P_n$ in parallel.
\par A first constant-depth circuit establishes the configuration -- which describes $P_n$ -- from $L_{DC}(P)$ and constant-depth circuits update this configuration after steps of normalization. Once the configuration of the normal form of $P_n$ is obtained, a last constant-depth circuit identifies the first proof net connected to the result tensor and establishes if it is $b_0$ or $b_1$ -- that is to say if the result of the evaluation is \textit{false} or \textit{true}.
\par As all the circuits are of constant depth, the depth of $C_n$ is linear in $d(P_n)$. The size of $C_n$ is $O(|P_n|^4)$: every circuit simulating a parallel reduction needs $O(|P_n|^3)$ gates and in the worst case -- if $d(P_n)$ is linear in the size of the proof circuit -- $O(|P_n|)$ steps are needed to normalize the proof net.
\par The \autoref{mBNPCC} helps us to conclude that $PCC^i \subseteq mBN^i \subseteq AC^i( \textit{UstConn}_2)$
\end{proo}
\par The simulation is slightly different from the translation: the Boolean circuit does not have to identify the pieces or any mechanism of computation of $P_n$, but simply to apply $\rightrightarrows_t$, $\rightrightarrows_a$, $\rightrightarrows_m$ to it until it reaches a normal form and then look at the value obtained. This simulation can be applied to any Boolean proof net, but in order to have results concerning complexity we preferred to stay in this uniform framework.
\begin{thm}
 For all $i \in \mathbb{N}$, $AC^i \subseteq PCC^i \subseteq AC^i(\textit{UstConn}_2)$.
\end{thm}
\begin{proo}
 By \autoref{NCtoBC} and \autoref{mbntoPN}. The key point is to notice -- as the reductions are in $AC^0$-- that $AC^0 \subseteq PCC^0 \subseteq AC^0(\textit{UstConn}_2)$.
\end{proo}
\begin{rem}
We focused on sublogarithmic classes of complexity, but we can draw more general conclusions by re-introducing $\textit{UstConn}_2$. From what precedes and from the fact that $\textit{UstConn}_2$ may be represented by Boolean proof nets of constant-depth and polynomial size, it is easy to conclude that for all $i \in \mathbb{N}$, $PCC^i(\textit{UstConn}_2) = AC^i(\textit{UstConn}_2) = mBN^i$.
\end{rem}
\section*{Conclusion}
\par By restricting ourselves to the uniform complexity classes and by lightening the simulation of the Boolean functions by proof nets, we established the validity of results given by \cite{terui04} and \cite{mogbil09} when extended to constant-depth Boolean circuits. Those complexity classes are of great interest as they are below $L$ and mostly used in reductions. This paper proves that proof nets for Multiplicative Linear Logic are a pertinent tool to study complexity classes, including very small ones, even if we do not use exponentials.
\par The simulation of the parallel elimination of $t$-cuts by Boolean circuits needs $\textit{UstConn}_2$ gates. But as $\textit{UstConn}_2 \in L$, there is for the time being no clue if a sublogarithmic Boolean circuit can simulate Boolean proof nets: $AC^0(\textit{UstConn}_2) \subseteq AC^1 \supseteq L$.
\par Our future work will aim to prove that proof nets are a model of computation as relevant as Alternating Turing Machines but easier to manipulate: as we are in an implicit complexity framework, the size of our object suffices to know in which class of complexity it rests, whereas the only way of knowing where is an ATM is to run it on inputs. We already have gateways -- by using correspondences with Boolean circuits --  between Boolean proof nets and ATM, but our objective is to establish direct proofs.
\vfill
\bibliographystyle{eptcs.bst}
\bibliography{aubert_biblio.bib}
\end{document}